\documentclass[letterpaper]{article} % DO NOT CHANGE THIS
\usepackage{aaai2027}  % DO NOT CHANGE THIS
\usepackage[hyphens]{url}  % DO NOT CHANGE THIS
\usepackage{graphicx} % DO NOT CHANGE THIS
\usepackage{natbib}  % DO NOT CHANGE THIS AND DO NOT ADD ANY OPTIONS TO IT
\usepackage{caption} % DO NOT CHANGE THIS AND DO NOT ADD ANY OPTIONS TO IT
\usepackage{algorithm}
\usepackage{algorithmic}
\usepackage{amsmath}
\usepackage{amssymb}
\usepackage{multirow} % 支持跨越多行的单元格（\multirow 命令）

\usepackage{newfloat}
\usepackage{listings}
\DeclareCaptionStyle{ruled}{labelfont=normalfont,labelsep=colon,strut=off} % DO NOT CHANGE THIS
\floatstyle{ruled}
\newfloat{listing}{tb}{lst}{}
\floatname{listing}{Listing}

\usepackage{booktabs}

\nocopyright %-- Your paper will not be published if you use this command
\title{GROW: Group-Relative Advantage-Weighted On-Policy Reinforcement Learning of Autoregressive-Diffusion Text-to-Speech Model}
\author{
    Guanrou Yang\textsuperscript{\rm 1,2},
    Tian Tan\textsuperscript{\rm 1},
    Qian Chen\textsuperscript{\rm 4},
    Ziyang Ma\textsuperscript{\rm 1,2},
    Yakun Song\textsuperscript{\rm 1,2},
    Zhikang Niu\textsuperscript{\rm 1,2},\\
    Qi Chen\textsuperscript{\rm 1,2},
    Wenming Tu\textsuperscript{\rm 1},
    Haitao Li\textsuperscript{\rm 2,5},
    Shan Yang\textsuperscript{\rm 3},
    Xie Chen\textsuperscript{\rm 1,2}\corresponding
}
\affiliations{
\textsuperscript{\rm 1}Shanghai Jiao Tong University
\textsuperscript{\rm 2}Shanghai Innovation Institute
\textsuperscript{\rm 3}Tencent
\textsuperscript{\rm 4}Independent Researcher
\textsuperscript{\rm 5}Zhejiang University
\{yangguanrou, chenxie95\}@sjtu.edu.cn
}

\begin{document}

\maketitle

\begin{abstract}
Reinforcement learning for flow-matching text-to-speech is complicated by deterministic ODE sampling: trajectory-level policy-gradient methods typically convert the ODE into an SDE and track per-step likelihood ratios, introducing stochastic perturbations and substantial overhead. 
We propose \textbf{GROW}, a group-relative advantage-weighted on-policy RL method that acts directly on the standard flow-matching objective. For each prompt, GROW samples a group of on-policy utterances, separately standardizes intelligibility and speaker-similarity rewards within the group, 
%combines them into a signed group-relative advantage, and uses this signal to reweight flow-matching regression. 
and combines them to reweight flow-matching regression. 
A Wasserstein-2 velocity penalty anchors the updated model to a frozen pretrained reference.
% We introduce a group-mean reward baseline to turn reward weighting into advantage weighting, 
% and further show that for strong pretrained TTS models with concentrated within-group rewards, positive exponential weights are dominated by a reward-agnostic self-imitation component, whereas the zero-mean signed advantage preserves the reward contrast required for effective credit assignment.
A group-mean reward baseline is introduced to convert reward weighting
into advantage weighting. For strong pretrained TTS models with
concentrated rewards, positive exponential weighting
is dominated by reward-agnostic self-imitation, whereas a zero-mean signed
advantage preserves effective within-group credit assignment.
Instantiated on DiTAR and evaluated on LibriSpeech and Seed-TTS EN/ZH, GROW
reduces average WER from $2.016$ to $1.558$ and raises speaker similarity
from $0.676$ to $0.715$ while keeping UTMOS.
%about $70\%$ and $30\%$ larger than those of our Flow-GRPO baseline. 
With 10-NFE training rollouts and 32-NFE evaluation, GROW retains comparable  performance while training $2.9\times$ faster than 32-NFE DiTAR-GRPO. %We will release our faithful DiTAR reproduction and complete GROW codebase.
We will open-source complete GROW codes, faithful DiTAR reproduction, and all model checkpoints.
\end{abstract}

\begin{links}
    \link{Code}{https://github.com/yanghaha0908/GROW}
    % \link{Datasets}{https://aaai.org/example/datasets}
    % \link{Extended version}{https://aaai.org/example/extended-version}
\end{links}

% Reinforcement learning (RL) for continuous flow-matching and
% autoregressive-diffusion text-to-speech (TTS) is difficult because deterministic
% ODE sampling exposes no tractable per-step transition probability for a policy
% gradient. 
% The prevailing remedy, exemplified by Flow-GRPO, rewrites the sampling
% ODE as an SDE, but this injects noise into the very utterances the reward models
% score and demands stochastic trajectories, per-step likelihood ratios, and
% reference-divergence tracking---costs that recur for every autoregressively
% generated patch. 
% We propose \textbf{GROW}, a concise online RL algorithm that
% instead acts directly through the ordinary flow-matching regression on clean,
% on-distribution samples. 

% Uncomment the following to link to your code, datasets, an extended version or similar.
% You must keep this block between (not within) the abstract and the main body of the paper.
% Make sure that you do not de-anonymize yourself with these links.
% \begin{links}
%     \link{Code}{https://aaai.org/example/code}
%     \link{Datasets}{https://aaai.org/example/datasets}
%     \link{Extended version}{https://aaai.org/example/extended-version}
% \end{links}

\section{Introduction}

Modern text-to-speech (TTS) systems can synthesize highly natural speech and
clone an unseen speaker from a short prompt. However, their pretraining objectives provide indirect supervision for deployment-relevant attributes, such as intelligibility and speaker identity.
This motivates task-aware post-training of autoregressive (AR) diffusion models, which combine fine-grained acoustic modeling in continuous latent spaces with causal AR modeling of long-range sequential dependencies.
% This motivates task-aware post-training for autoregressive diffusion models, which preserve fine acoustic details of continuous representations while retaining the sequential modeling capacity of autoregressive generation.
% Specifically, DiTAR~\citep{jia2025ditar} models dependencies across continuous
% latent patches with a causal language model(LM) and generates frames within each
% patch with a diffusion head. This division avoids discrete codec
% quantization and supports strong zero-shot synthesis, but it also presents a
% post-training problem that differs from both token-level LM
% alignment and non-autoregressive diffusion alignment, for an
% utterance is produced by repeatedly solving a deterministic flow ODE inside an
% autoregressive loop, making conventional likelihood-based reinforcement
% learning (RL) tricky and expensive.
Specifically, DiTAR~\citep{jia2025ditar} uses a causal language model (LM) to capture dependencies across continuous latent patches and a diffusion head to generate frames within each patch. This hybrid design avoids discrete codec quantization and enables strong zero-shot synthesis. However, it also creates a distinct post-training challenge: generating an utterance requires repeatedly solving a deterministic flow ODE within an autoregressive loop, making likelihood-based reinforcement learning (RL) substantially more difficult and expensive than alignment for discrete-token LMs or NAR diffusion models.

RL-based post-training has become increasingly important for
speech generation. Most prior efforts act on discrete codec LMs. Seed-TTS compares REINFORCE, PPO, and DPO under rewards for recognition,
speaker similarity, and emotion, while DiffRO makes token-level rewards
differentiable through Gumbel--Softmax and a learned multi-task reward
model~\citep{Seed-tts,DiffRO,Cosyvoice3}. GLM-TTS employs GRPO with
multiple rewards~\citep{Glm-tts}, and a recent empirical
study compares and combines GRPO with DiffRO in an audio LM
training framework~\citep{gao2025explore}. These studies demonstrate the benefits
of post-training for codec-LM TTS.
%but do not address continuous flow-matching or diffusion speech generators.

Alignment of continuous generators has followed several routes.
ARDM-DPO fine-tunes DiTAR from offline preference pairs, while VGPO learns a
causal value model whose gradient updates the model~\citep{liu2026direct,liu2026vgpo}. F5R-TTS constructs a
probabilistic flow-matching TTS model for GRPO, and FlowTTS-GRPO introduces Flow-GRPO to TTS~\citep{F5r-tts,flowtts-grpo,flowgrpo}. %by converting ODE paths to SDE
% FlowTTS-GRPO makes existing flow samplers stochastic before applying trajectory-level policy gradients 
The latter route also extends to
text-to-audio generation in Resonate~\citep{resonate}. 
% A complementary family 
A complementary line of work, developed largely for image, 
directly incorporates reward signals into the forward flow-matching
regression used for pretraining. 
ORW-CFM-W2 applies exponential reward weighting with Wasserstein-2 regularization.
DiffusionNFT contrasts positive and negative synthesis to derive an implicit policy-improvement direction.
% AWM reweights the standard score/flow-matching loss with group-relative advantages.
AWM retains the standard score/flow-matching objective used in pretraining and reweights each sample’s loss by its advantage. 
FlowAWR regresses towards an advantage-rectified velocity field
~\citep{orw,diffusionnft,AWM,flowawr}. 
This forward-process paradigm begins to transfer to speech. SwanVoice
applies DiffusionNFT post-training to its flow-matching generator, while
dots.tts applies the reward-free SOAR self-correction strategy to the
flow-matching head of a continuous AR model
~\citep{swanvoice,dotstts}. 

%Our goal is a simple online method for flow-matching TTS that can use black-box utterance-level feedback while retaining deterministic sampling and the standard training pipeline.

% A widely used Flow-GRPO recipe exposes the central difficulty. Because a
% flow-matching sampler is deterministic, Flow-GRPO converts its sampling ODE into
% an SDE so that every denoising transition has a tractable Gaussian density and
% can enter a group-relative policy gradient~\citep{flowgrpo}. This construction
% is general, but it introduces stochastic perturbations into the utterances
% scored by the reward models, stores and replays sampling trajectories, and
% tracks likelihood ratios and reference divergence step by step. In DiTAR, these
% costs recur for every autoregressively generated patch. Mixed ODE--SDE windows
% and coefficients-preserving sampling mitigate the overhead and accumulated
% noise~\citep{li2025mixgrpo,wang2025cps}, but they do not remove the stochastic
% sampler or the trajectory-level policy-gradient machinery. The resulting
% training procedure remains substantially more involved than the regression
% loss used to pretrain the flow model.
The two paradigms address the likelihood bottleneck differently. Flow-GRPO
makes policy gradients tractable by converting the sampling ODE into an SDE,
but requires trajectory storage and per-step computation of policy likelihood ratios and divergences to the frozen reference, costs that recur for
every generated patch in DiTAR. MixGRPO and
CPS reduce the overhead and accumulated noise, yet retain stochastic sampling
and trajectory-level optimization~\citep{li2025mixgrpo,wang2025cps}.
Direct reward-weighted flow-matching regression avoids this machinery, but positive
exponential weighting poses a different challenge for strong pretrained TTS
models. Bounded recognition and speaker-similarity rewards are often tightly
clustered within a prompt group, causing a shared self-imitation component to dominate the weak reward contrast. This regime calls for a weighting scheme that amplifies subtle within-prompt differences while accommodating heterogeneous reward scales.

% Two recent studies motivate our approach from complementary perspectives.
% ORW-CFM-W2~\citep{orw} established likelihood-free online flow
% fine-tuning through nonnegative reward-weighted regression and introduced a
% Wasserstein-2 regularizer; AWR~\citep{AWR} incorporated group-relative
% advantages directly into score/flow-matching training. Inspired by this line of
% work, we investigate direct reward-weighted flow-matching for DiTAR and identify
% a challenge specific to strong pretrained TTS models. Recognition and speaker
% similarity rewards are bounded and often tightly clustered among candidates for
% the same prompt. Under positive exponential weighting, all candidates are
% therefore reinforced by a large shared component, while the reward-dependent
% contrast that drives improvement remains weak. GROW instead turns these subtle
% within-prompt differences into a signed, scale-normalized learning signal and
% balances multiple TTS objectives with different reward distributions.

We introduce \textbf{GROW}, a group-relative, advantage-weighted
on-policy RL algorithm for flow-matching TTS. We apply it to DiTAR, an
AR diffusion TTS model, but it can also be applied to conventional flow-matching TTS models such as F5-TTS~\citep{chen2025f5}.
For each prompt, GROW samples a group of utterances from the current policy and scores them for intelligibility
and speaker similarity. It normalizes each reward within the group, combines
them into a signed group-relative advantage, and uses this advantage to reweight
the standard flow-matching loss. A Wasserstein-2 distance loss anchors the updated
velocity field to the frozen pretrained model. 
% The update requires only a
% flow-matching training objective and is therefore not specific to
% autoregressive generation; in principle, it can also be applied to conventional
% flow-matching TTS models such as F5-TTS~\citep{chen2024f5tts}, although we
% evaluate it only on DiTAR. 
This simple procedure preserves deterministic ODE
sampling and requires neither ODE-to-SDE conversion nor per-step likelihood
evaluation.
On LibriSpeech-PC test-clean and the English and Chinese Seed-TTS test sets, GROW improves
both target criteria while maintaining UTMOS at the pretrained level. Averaged over the three
test sets, it reduces WER from $2.016$ to $1.558$ with a $0.458$ absolute reduction
and raises speaker similarity from $0.676$ to $0.715$ with a $0.039$ absolute gain, surpassing those of DiTAR-GRPO baseline.
% These improvements are approximately $70\%$ and $30\%$ larger, respectively,
% than those of our in-house DiTAR Flow-GRPO baseline.
% strengthened with MixGRPO and CPS. 
GROW is also training-efficient. With 10-NFE rollouts, it achieves comparable performance to its 32-NFE setting while training $2.9\times$ faster than the 32-NFE DiTAR-GRPO baseline. All models are evaluated with 32 NFE. At matched training-rollout NFE, GROW trains about $1.1\times$ faster than DiTAR-GRPO.
% We will release our faithful reproduction of DiTAR pretraining and the complete GROW reinforcement-learning implementation.
Our contributions are summarized as follows:
\begin{itemize}
    % \item We propose GROW, a concise and effective online RL algorithm that
    % reweights the ordinary flow-matching regression by a
    % signed group-relative advantage under Wasserstein-2
    % reference anchor. Inheriting the likelihood-free, SDE-free, clean-sample advantages of forward-process flow RL,
    % it applies to any flow-matching generator.
    \item We propose a concise and effective on-policy RL method
    that directly applies signed group-relative advantage weighting to
    flow-matching regression with a Wasserstein-2 reference anchor.
    % \item We analyze why positive exponential weighting becomes dominated by a
    % common self-imitation component when pretrained TTS rewards are concentrated,
    % and instantiate a signed, group-normalized advantage for multi-objective TTS, showing substantial gains over exponential weighting of either raw
    % rewards or group-relative advantages.
\item We introduce a group-mean reward baseline to turn reward weighting into advantage weighting, then replace positive exponential weights with a signed, group-normalized advantage for effective credit assignment.
% \item We introduce a group-mean reward baseline to convert reward
% weighting into advantage weighting, then replace positive exponential
% weights with a signed, group-normalized advantage for within-group
% credit assignment.
% Weighting ablations 
 \item We demonstrate consistent gains and superior training efficiency over Flow-GRPO on three
    multilingual zero-shot benchmarks and validate the effects of rollout NFE, CFG, reward
    weighting and composition.
    
\end{itemize}

% We also plan to release cleaned DiTAR pretraining and GROW fine-tuning code.

% Uncomment the following to link to your code, datasets, an extended version or similar.
% You must keep this block between (not within) the abstract and the main body of the paper.
% Make sure that you do not de-anonymize yourself with these links.
% \begin{links}
%     \link{Code}{https://aaai.org/example/code}
%     \link{Datasets}{https://aaai.org/example/datasets}
%     \link{Extended version}{https://aaai.org/example/extended-version}
% \end{links}

% \section{Related Work}

% \section{Methodology}

%GROW: Online Reward-Weighted Flow-Matching RL for Autoregressive TTS with Wasserstein-2 Regularization
\begin{algorithm*}[t]
\caption{GROW: Group-Relative Advantage-Weighted On-Policy RL of AR-Diffusion TTS model}
\label{alg:grow}
\textbf{Input}: pretrained DiTAR $\theta_{\mathrm{pre}}$; dataset
$\mathcal{D}=\{(x^{\mathrm{ref}},y^{\mathrm{ref}},y)\}$ of (prompt wav,
prompt text, target text); VAE encoder $\mathcal{E}$.
% Speaker model $\mathrm{SPK}(\cdot)$, and ASR model $\mathrm{ASR}(\cdot)$.\\
\textbf{Parameter}: group size $G$; W2 weight $\beta_{W2}$; reward weights
$\lambda_{\mathrm{sim}},\lambda_{\mathrm{wer}}$; learning rate $\eta$.
\textbf{Output}: fine-tuned DiTAR $\theta$.
\begin{algorithmic}[1]
\STATE $\theta\leftarrow\theta_{\mathrm{pre}}$;\quad $\theta_{\mathrm{ref}}\leftarrow\theta_{\mathrm{pre}}$ (frozen)
\WHILE{not converged}
  \STATE draw $(x^{\mathrm{ref}},y^{\mathrm{ref}},y)\sim\mathcal{D}$, roll out $G$ candidates $\hat x^{(k)}\sim\pi_\theta\big(\cdot\mid x^{\mathrm{ref}},y^{\mathrm{ref}},y\big)$, $k=1,\ldots,G$
  % with independent sampling noise $\epsilon^{(k)}\!\sim\!\mathcal{N}(0,I)$, 
  \STATE compute rewards: $R^{(k)}_{\mathrm{sim}}\leftarrow\cos\!\big(\mathrm{SPK}(\hat x^{(k)}),\mathrm{SPK}(x^{\mathrm{ref}})\big)$;\quad $R^{(k)}_{\mathrm{wer}}\leftarrow 1-\mathrm{WER}\!\big(\mathrm{ASR}(\hat x^{(k)}),\,y\big)$
% \STATE $z^{(k)}_r\leftarrow\dfrac{R^{(k)}_r-\mu_r}{\sigma_r+\varepsilon}$ \ for $r\in\{\mathrm{sim},\mathrm{wer}\}$, \ where $\mu_r,\sigma_r$ are the group mean and std of $\{R^{(k)}_r\}_{k=1}^{G}$
\STATE compute weights: $z^{(k)}_r\leftarrow
\bigl(R^{(k)}_r-\mu_r\bigr)\big/\bigl(\sigma_r+\varepsilon\bigr)$
for $r\in\{\mathrm{sim},\mathrm{wer}\}$;\quad $w^{(k)}\leftarrow\lambda_{\mathrm{wer}}\,z^{(k)}_{\mathrm{wer}}+\lambda_{\mathrm{sim}}\,z^{(k)}_{\mathrm{sim}}$ %$\tilde y\leftarrow[\,y^{\mathrm{ref}}\Vert y]$
  \STATE assemble training data: \ $u^{(k)}\leftarrow[\,x^{\mathrm{ref}}\Vert\hat x^{(k)}]$; \ \ $x_1^{(k)}\leftarrow\mathcal{E}(u^{(k)})$
  % \STATE sample $t\sim U[0,1]$, \ $x_0^{(k)}\sim\mathcal{N}(0,I)$;\quad $x_t^{(k)}\leftarrow(1-t)\,x_0^{(k)}+t\,x_1^{(k)}$;\quad $v^{\star(k)}\leftarrow x_1^{(k)}-x_0^{(k)}$
  \STATE sample $t^{(k)}\sim U[0,1]$, \ $x_0^{(k)}\sim\mathcal{N}(0,I)$;\quad $x_t^{(k)}\leftarrow(1-t^{(k)})\,x_0^{(k)}+t^{(k)}\,x_1^{(k)}$;\quad $v^{\star(k)}\leftarrow x_1^{(k)}-x_0^{(k)}$
  % \STATE compute loss: $\mathcal{L}\leftarrow\big\langle\,w^{(k)}\,\big\lVert v_\theta(x_t^{(k)},t)-v^{\star(k)}\big\rVert_D^2\,\big\rangle+\beta\,\big\langle\,\big\lVert v_\theta(x_t^{(k)},t)-v_{\theta_{\mathrm{ref}}}  (x_t^{(k)},t)\big\rVert_D^2\,\big\rangle$
  \STATE compute loss: $\mathcal{L}\leftarrow \big\langle\,w^{(k)}\,\big\lVert v_\theta(x_t^{(k)},t^{(k)})-v^{\star(k)} \big\rVert_D^2\,\big\rangle +\beta_{\mathrm{W2}}\,\big\langle\,\big\lVert v_\theta(x_t^{(k)},t^{(k)}) -v_{\theta_{\mathrm{ref}}}(x_t^{(k)},t^{(k)}) \big\rVert_D^2\,\big\rangle$

  % \STATE $\mathcal{L}\leftarrow\big\langle\,w^{(k)}\,\big\lVert v_\theta(x_t^{(k)},t,c^{(k)})-v^{\star(k)}\big\rVert_D^2\,\big\rangle+\beta\,\big\langle\,\big\lVert v_\theta(x_t^{(k)},t,c^{(k)})-v_{\theta_{\mathrm{ref}}}(x_t^{(k)},t,c^{(k)})\big\rVert_D^2\,\big\rangle$
  \STATE update: $\theta\leftarrow\mathrm{AdamW}\big(\theta,\ \nabla_\theta\mathcal{L},\ \eta\big)$
\ENDWHILE; \textbf{return} $\theta$
% \STATE 
\end{algorithmic}
\end{algorithm*}

\section{Preliminaries}

% \subsection{DiTAR: Autoregressive Modeling of Continuous Speech Latents}
\subsection{DiTAR: Autoregressive-Diffusion TTS}
We adopt DiTAR, an AR diffusion model that
couples a causal LM with a diffusion Transformer to generate
speech in a continuous latent space, avoiding the information loss of
discrete acoustic tokens. A variational auto-encoder first maps a waveform to a
sequence of latent frames $x=(x_1,\dots,x_N)$ with $x_i\in\mathbb{R}^{D}$. 
% The VAE is frozen throughout. 
% Rather than predicting one frame at a time, 
DiTAR partitions the sequence into patches of $P$ consecutive frames.
% and follows a divide-and-conquer factorization: 
A causal Transformer models the dependencies across patches, while a diffusion head models the frames within
each patch. Writing $\theta=(\theta_a,\theta_b)$ for the two components,
% \begin{equation}
% p_{\theta}(x)=\prod_{i} \underbrace{p_{\theta_a}\!\big(h_i\mid x_{\le i}\big)}_{\text{inter-patch (causal LM)}}\;
% \underbrace{p_{\theta_b}\!\big(x_{i+1:i+P}\mid h_i\big)}_{\text{intra-patch (diffusion)}},
% \label{eq:ditar-factor}
% \end{equation}
\begin{equation}
p_{\theta}(x)
=\prod_i
p_{\theta_a}\!\left(h_i\mid x_{\le i}\right)
p_{\theta_b}\!\left(x_{i+1:i+P}\mid h_i\right),
\label{eq:ditar-factor}
\end{equation}
where an aggregation encoder compresses each patch into a single embedding for
the LM, and $h_i$ is the LM state that summarizes the
history $x_{\le i}$ and conditions the diffusion head on the next patch.

The intra-patch generator is a bidirectional \emph{Local Diffusion Transformer}
(LocDiT). Given the context $h_i$, a time embedding, and previously
generated patches as prefix, LocDiT denoises the next patch under a linear
flow-matching objective: with noise
$x_0\sim\mathcal{N}(0,I)$, clean latent $x_1$, and $t\sim U[0,1]$,
% \begin{equation}
% \begin{aligned}
% x_t&=(1-t)\,x_0+t\,x_1, \mathcal{L}_{\mathrm{FM}}(\theta)&=\mathbb{E}\big[\lVert v_\theta(x_t,t)-(x_1-x_0)\rVert^2\big],
% \end{aligned}
% \label{eq:cfm}
% \end{equation} 
% \begin{equation}
% x_t=(1-t)x_0+tx_1, 
% \mathcal{L}_{\mathrm{FM}}(\theta)
% =\mathbb{E}\!\left\lVert
% v_\theta(x_t,t)-(x_1-x_0)
% \right\rVert_2^2 .
% \label{eq:cfm}
% \end{equation}
% \begin{equation}
% \resizebox{\columnwidth}{!}{
% x_t=(1-t)x_0+tx_1, 
% \mathcal{L}_{\mathrm{FM}}
% =\mathbb{E}\bigl\|v_\theta(x_t,t)-(x_1-x_0)\bigr\|_2^2 .
% \label{eq:cfm}
% }
% \end{equation}
\begin{equation}
\resizebox{\dimexpr\columnwidth-2.5em\relax}{!}{$\displaystyle
x_t=(1-t)x_0+tx_1,\quad
\mathcal{L}_{\mathrm{FM}}(\theta)
=\mathbb{E}\bigl[
\lVert v_\theta(x_t,t)-(x_1-x_0)\rVert_2^2
\bigr]
$}
\label{eq:cfm}
\end{equation}
% so the network regresses the target velocity $x_1-x_0$ from a conditioning $c$
% that comprises the language-model state $h_i$ and the prefix patches. 
At inference the model runs
autoregressively over patches: for each patch it solves the probability-flow ODE, 
% with a number of steps
decodes sampled latents, and a stop head
predicts the end of the utterance. 
% Zero-shot voice cloning is cast as conditional
% continuation---the prompt speech, prompt transcript, and target text are placed
% in the prefix, and the model continues speaking in the prompt's voice.
Classifier-free guidance (CFG) is applied by intermittently dropping the conditioning
during training and interpolating conditional and unconditional velocities at
sampling time.

\subsection{RL for Flow-Matching Models}
\label{sec:prelim-rl}
% We fine-tune DiTAR so that its outputs score highly under task rewards while
% staying close to the pretrained model. TODO rl的目标是xxx
% RL post-training aims to steer a generative model toward outputs that better satisfy task-specific objectives.
Let $\pi_{\theta}$ denote the generative
policy induced by the model over complete utterances, and let $q$ be the
distribution of the frozen pretrained model, also serving as reference.
For a scalar reward $r(\cdot)$, the reward-maximization objective with a
Kullback--Leibler penalty toward the reference is
\begin{equation}
\max_{\pi_{\theta}}\ \; \mathbb{E}_{x\sim\pi_{\theta}}\big[r(x)\big]
\;-\;\tfrac{1}{\tau}\,\mathrm{KL}\!\left(\pi_{\theta}\,\Vert\,q\right),
\label{eq:rl-obj}
\end{equation}
where $\tau>0$ is an inverse temperature trading reward against deviation from the
reference. This objective has the well-known closed-form optimum
\begin{equation}
\pi^{\star}(x)\;\propto\;q(x)\,\exp\!\big(\tau\,r(x)\big),
\label{eq:opt-policy}
\end{equation}
that is, the reward-optimal distribution is the reference distribution
exponentially tilted toward high reward. $\tau\!\to\!0$ recovers the pretrained
model and larger $\tau$ concentrates mass on high-reward
samples~\citep{peters2010relative}.

For flow-matching generators, the exponential tilt in Eq.~\eqref{eq:opt-policy}
can be realized through reward-weighted regression (RWR) without explicitly
evaluating the model likelihood. 
Given on-policy samples $x\sim\pi_n$, reward-weighted flow matching
scales each sample's standard regression loss by
$w(x)=\exp(\tau r(x))$. Under the idealized
assumption that the model perfectly fits the weighted flow-matching objective each round, the updated policy satisfies
$\pi_{n+1}(x)\propto\pi_n(x)\exp\bigl(\tau r(x)\bigr)$
\citep{orw,peters2007reinforcement}.

Thus, rewards can act directly through the original flow-matching objective without % stochasticizing the sampling ODE or computing trajectory likelihood ratios. 
ODE-to-SDE conversion or trajectory likelihood evaluation.
GROW builds on this direct-regression view, but revisits how the rollout rewards should be
converted into regression weights for a strong pretrained TTS model.

\section{GROW}
\label{sec:method}
GROW fine-tunes pretrained DiTAR by applying RL directly to its
flow-matching acoustic head. Its design combines on-policy
multi-objective rewards, a signed group-normalized advantage that
replaces positive exponential weighting, and Wasserstein-2 distance anchoring
to the frozen pretrained model. Algorithm~\ref{alg:grow} summarizes
the full procedure, and the following subsections detail these
components.
% GROW fine-tunes pretrained DiTAR with RL
% applied directly to its flow-matching acoustic head. For a given prompt, it rolls
% out a group of on-policy candidates, scores each for speaker similarity
% and intelligibility, and turns these scores into a signed, group-relative
% advantage that reweights flow-matching loss on the generated frames, while a
% Wasserstein-2 velocity penalty anchors the update to the frozen pretrained model.
% We detail each component below. Algorithm~\ref{alg:grow} summarizes the full procedure.

\subsection{On-Policy Rollouts and Multi-Objective Rewards}
Each optimization step draws one example $(x^{\mathrm{ref}},y^{\mathrm{ref}},y)$
---prompt waveform, its transcript, and target text---and generates a group of
$G$ candidates $\{\hat x^{(k)}\}_{k=1}^{G}$ from the current policy.
%conditioned on the prompt and the target text. 
Each rollout uses different sampling noise to form diverse continuations of the same prompt. Since each rollout is sampled from the current model, training remains
on-policy and the data distribution evolves with the policy throughout
optimization. Each candidate is
scored on two axes that matter for zero-shot TTS, speaker similarity to the
prompt and intelligibility of the target text, using
\mbox{$R_{\mathrm{sim}}^{(k)}
=\cos\!\bigl(\mathrm{SPK}(\hat{x}^{(k)}),
\mathrm{SPK}(x^{\mathrm{ref}})\bigr)$}
and
\mbox{$R_{\mathrm{wer}}^{(k)}
=1-\mathrm{WER}\!\bigl(\mathrm{ASR}(\hat{x}^{(k)}),y\bigr)$}, where $\mathrm{SPK}$ is a speaker encoder and $\mathrm{ASR}$ a speech recognizer.
Both reward models are frozen.% and used without gradients.
% $R_{\mathrm{sim}}^{(k)}
% =\cos\!\bigl(\mathrm{SPK}(\hat x^{(k)}),\mathrm{SPK}(x^{\mathrm{ref}})\bigr)$
% and
% $R_{\mathrm{wer}}^{(k)}
% =1-\mathrm{WER}\!\bigl(\mathrm{ASR}(\hat x^{(k)}),y\bigr)$
% \begin{align}
% R^{(k)}_{\mathrm{sim}}&=\cos\!\big(\mathrm{SPK}(\hat x^{(k)}),\,\mathrm{SPK}(x^{\mathrm{ref}})\big),
% \label{eq:r-sim}\\
% R^{(k)}_{\mathrm{wer}}&=1-\mathrm{WER}\!\big(\mathrm{ASR}(\hat x^{(k)}),\,y\big),
% \label{eq:r-wer}
% \end{align}

\subsection{From Reward Weighting to Advantage Weighting}
\label{sec:reward-to-adv}
% In policy-gradient RL, a baseline is an action-independent control
% variate subtracted from the sampled return. Its expected contribution
% to the policy gradient is zero, while an appropriate baseline can
% substantially reduce estimator variance
% \citep{williams1992simple}. %,sutton2000policy
In policy-gradient RL, subtracting a baseline that is independent of
the sampled action leaves the expected policy gradient unchanged,
while an appropriate baseline can reduce the variance of its
Monte Carlo estimate~\citep{williams1992simple,sutton1999policy}.
Scoring each generated sample by the exponential of its own reward,
$w(x)=\exp(\tau r(x))$, is the classical RWR
form, which realizes the tilted optimum~\eqref{eq:opt-policy}. 
To adapt to RL fine-tuning of a pretrained TTS model, we begin by introducing a reward baseline.
Centering the reward by a constant baseline, $r(x)\!\to\!r(x)-b$, leaves the
tilted optimum~\eqref{eq:opt-policy} unchanged, for the constant factor $e^{-\tau b}$
is absorbed into the normalizer. %, while reducing the variance of the finite-sample gradient estimate~\citep{AWR}.  % without introducing bias TODO主要是这句
A natural and widely used baseline is
the expected return under the current policy, which in group-relative methods is
estimated by the group mean~\citep{shao2024deepseekmath}. In our
bandit-style setting, where a prompt is a state, and each rollout is an action, the
expected return under the current policy is naturally estimated by the group mean
$\bar R=\tfrac1G\sum_k R^{(k)}$. Centering the reward therefore turns the
reward-weighted form into the advantage-weighted form
\begin{equation}
w^{(k)}=\exp\!\big(\tau\,A^{(k)}\big),\qquad A^{(k)}=R^{(k)}-\bar R,
\label{eq:exp-adv}
\end{equation}
% which shares the optimum of \eqref{eq:opt-policy} but 
removing the reward-scale factor $\exp(\tau\bar R)$ that otherwise multiplies the update.

% \subsection{From Reward Weighting to Advantage Weighting}
% \label{sec:reward-to-adv}

% In policy-gradient RL, subtracting a baseline that is independent of
% the sampled action leaves the expected policy gradient unchanged,
% while an appropriate baseline can reduce the variance of its
% Monte Carlo estimate~\citep{williams1992simple,sutton2000policy}.
% Choosing the state value as the baseline yields an advantage, which
% measures an action's return relative to the policy's expected return
% at the same state.

% Classical RWR assigns each generated sample the exponential weight
% $w(x)=\exp(\tau r(x))$, which realizes the tilted optimum in
% Eq.~\eqref{eq:opt-policy}. For any baseline $b$ shared across
% candidates for the same prompt,
% $\exp\!\bigl(\tau(r(x)-b)\bigr)
% =\exp(-\tau b)\exp(\tau r(x))$.
% The common factor $\exp(-\tau b)$ is absorbed into the per-prompt
% normalizer, leaving the reward-tilted distribution unchanged.

% In our bandit-style setting, a prompt serves as the state and each
% rollout as an action. Following group-relative methods, we estimate
% the prompt-conditioned expected reward using the group mean
% $\bar R=\tfrac{1}{G}\sum_k R^{(k)}$
% \citep{shao2024deepseekmath}. Centering by this group baseline yields
% the advantage-weighted form
% \begin{equation}
% w^{(k)}=\exp\!\bigl(\tau A^{(k)}\bigr),
% \qquad
% A^{(k)}=R^{(k)}-\bar R,
% \label{eq:exp-adv}
% \end{equation}
% which preserves the same per-prompt tilted distribution while removing
% the common factor $\exp(\tau\bar R)$ from the unnormalized regression
% weights.

\subsection{From Exponential to Linear Group-Relative Advantage}
\label{sec:exp-to-linear}
We now specialize \eqref{eq:exp-adv} to the reward statistics of TTS. 
% Since the policy starts from a strong pretrained model, and both rewards are bounded and
% near-saturated, with speaker similarity and WER both near-perfect
% for a competent model, the within-group rewards are highly concentrated, so the
% advantages $A^{(k)}$ are small in magnitude. 
Since the strong pretrained policy already achieves high speaker
similarity and low WER, both bounded rewards are near saturation and
highly concentrated, yielding small advantages $A^{(k)}$.
A first-order Taylor expansion of the
exponential weight about $A^{(k)}=0$,
\begin{equation}
\exp\!\big(\tau A^{(k)}\big)=1+\tau A^{(k)}+O\!\big((\tau A^{(k)})^2\big),
\label{eq:taylor}
\end{equation}
separates the weight into a constant term of $1$ and a linear term $\tau
A^{(k)}$. 
%To see why the constant term carries no learning signal, 
Write the
advantage-weighted loss over the group as $\mathcal{L}_{\mathrm{aw}}=\sum_{k}
w^{(k)}\ell^{(k)}$, where $\ell^{(k)}$ is the flow-matching loss of rollout $k$.
Decomposing the weight into its group mean and deviation,
$w^{(k)}=\bar w+\delta^{(k)}$ with $\sum_k\delta^{(k)}=0$,
\begin{equation}
\mathcal{L}_{\mathrm{aw}}=\underbrace{\bar w\,{\textstyle\sum_k}\ell^{(k)}}_{\text{self-imitation}}
\;+\;\underbrace{{\textstyle\sum_k}\delta^{(k)}\ell^{(k)}}_{\text{reward-contrast}}.
\label{eq:decomp}
\end{equation}
% \tfrac1G
% The first term is ordinary flow matching on the model's own rollouts, which is behavior cloning of the current policy. Since $\bar w$ is the same constant for every rollout, this term weights all candidates equally regardless of their rewards; it
% does no within-group credit assignment and merely reinforces the current policy. 
The first term reduces to ordinary flow matching on the policy's own
rollouts, which is self-imitation. Since the shared weight $\bar w$
ignores reward differences, it provides no within-group credit
assignment and merely reinforces the current policy.
Only the reward-contrast term moves probability mass, pulling toward
above-average rollouts and, when the weight is signed, pushing away from
below-average ones. For a strictly positive weight such as $\exp(\tau A)$, the
mean $\bar w$ stays of order one while the contrast deviations scale as
$\delta^{(k)}\!\propto\!\tau A^{(k)}$. The ratio of reward signal to
self-imitation is therefore of order $\tau\sigma$, where $\sigma$ is the
within-group advantage standard deviation. In the concentrated TTS regime, $\sigma$
is small, so unless $\tau$ is enlarged, the update is dominated by self-imitation
and provides little reward signal. Enlarging $\tau$ is not a robust fix, for
the useful magnitude scales with the unknown and drifting $\sigma$, so a fixed
temperature does not transfer across different prompts or training steps.

These observations point to a weight that is centered, signed, and normalized by
the group scale---the linear group-relative advantage,
% \begin{equation}
% w^{(k)}=\frac{A^{(k)}}{\sigma}=\frac{R^{(k)}-\bar R}{\operatorname{std}(R)+\varepsilon}.
% \label{eq:linear-adv}
% \end{equation}
\begin{equation}
w^{(k)}
= A^{(k)}\big/\sigma
= \bigl(R^{(k)}-\bar R\bigr)\big/\bigl(\operatorname{std}(R)+\varepsilon\bigr).
\label{eq:linear-adv}
\end{equation}
The linear advantage weight $A^{(k)}/\sigma$ is the leading-order direction of
the exponential advantage weight~\eqref{eq:exp-adv} when the within-group
advantages are small. It is obtained from $\exp(\tau A^{(k)})$ by four steps:
(i) replacing the exponential by the leading, linear term $1+\tau A^{(k)}$ of its
Taylor expansion; (ii) dropping the self-imitation constant $1$; (iii) folding the
positive scalar $\tau$ into the learning rate; and (iv) normalizing by the group
standard deviation $\sigma$. 
% This correspondence is a policy-gradient motivation
% rather than an exact identity. The equivalence between an exponential weight and
% its linear term holds only for the log-likelihood surrogate used in policy
% gradients. Our objective is instead a squared flow-matching regression, for which this equivalence does not carry over.
The resulting weight adopts the same group-normalized advantage estimator of GRPO, and is applied directly to flow-matching regression. It is zero-mean, so
its self-imitation component vanishes and the update is almost purely
reward-contrast. Its signed values pull the model toward
above-average candidates and away from below-average ones. Its normalization
makes the weight scale independent of the raw reward range, avoiding the need
to tune $\tau$ against the unknown and drifting $\sigma$.

The same normalization makes multi-objective rewards easy to combine. Speaker
similarity and WER have different scales and variances, so a weighted
sum of raw rewards would let the higher-variance stream dominate the advantage.
We therefore standardize each stream within the group and then fuse them,
% \begin{equation}
% \begin{aligned}
% z^{(k)}_{r}&=\frac{R^{(k)}_{r}-\mu_{r}}{\sigma_{r}+\varepsilon},\quad r\in\{\mathrm{sim},\mathrm{wer}\},\\
% w^{(k)}&=\lambda_{\mathrm{wer}}\,z^{(k)}_{\mathrm{wer}}+\lambda_{\mathrm{sim}}\,z^{(k)}_{\mathrm{sim}},
% \end{aligned}
% \label{eq:advantage}
% \end{equation}
  \begin{equation}
  w^{(k)}=\!\!\sum_{r\in\{\mathrm{sim},\mathrm{wer}\}}\!\!\lambda_{r}\,z^{(k)}_{r},\quad
  z^{(k)}_{r}=\frac{R^{(k)}_{r}-\mu_{r}}{\sigma_{r}+\varepsilon}.
  \label{eq:advantage}
  \end{equation}
where $\mu_r,\sigma_r$ are the group mean
and standard deviation of $\{R^{(k)}_r\}_{k=1}^{G}$. 
% Because each stream is
% rescaled to unit variance first, the two objectives contribute in the intended
% proportion regardless of their raw scales, in line with the multi-objective
% reward fusion used in recent flow-matching TTS RL~\citep{sun2025f5r,wang2026flowttsgrpo}. The resulting $w^{(k)}$ is a signed, zero-mean advantage. 
Following the dynamic-sampling strategy of DAPO~\citep{yu2026dapo}, we
discard and redraw any group with zero reward variance
($\sigma_{\mathrm{sim}}\!+\!\sigma_{\mathrm{wer}}=0$), which carries no gradient.

  % 备选 A（保留展开式，但把 $r$ 说明移到文字里，两式并排一行）：
  % \begin{equation}
  % z^{(k)}_{r}=\frac{R^{(k)}_{r}-\mu_{r}}{\sigma_{r}+\varepsilon},\quad
  % w^{(k)}=\lambda_{\mathrm{wer}}z^{(k)}_{\mathrm{wer}}+\lambda_{\mathrm{sim}}z^{(k)}_{\mathrm{sim}}.
  % \label{eq:advantage}
  % \end{equation}
  % （把 $r\in{\mathrm{sim},\mathrm{wer}}$ 挪到正文说明；若仍偏宽，去掉 \, 薄空格。）

\subsection{Advantage-Weighted Flow Matching with Reference Anchoring}
\label{sec:objective}
Each rollout is reassembled into a training example for the loss. 
The waveform
$u^{(k)}=[\,x^{\mathrm{ref}}\Vert\hat x^{(k)}]$ prepends the prompt to the
generated audio. 
%and the text $\tilde y=[\,y^{\mathrm{ref}}\Vert y]$ prepends the prompt transcript to the target text. 
Extracting VAE of waveform gives the target latents $x_1^{(k)}=\mathcal{E}(u^{(k)})$.  We then sample
a flow-matching time $t^{(k)}\sim U[0,1]$ and noise $x_0^{(k)}\sim\mathcal{N}(0,I)$
per rollout, and form the interpolant $x_t^{(k)}$ at time $t^{(k)}$,
\begin{equation}
x_t^{(k)}=(1-t^{(k)})\,x_0^{(k)}+t^{(k)}\,x_1^{(k)}, v^{\star(k)}=x_1^{(k)}-x_0^{(k)}.
\label{eq:interp}
\end{equation}
% The policy and the frozen reference evaluate their velocity fields on the same
% noised input $x_t^{(k)}$, turning the regularizer into a velocity discrepancy, following~\citep{orw}. 
% Evaluated on the same noised input $x_t^{(k)}$, this regularizer reduces to a velocity discrepancy between policy and reference, following~\citep{orw}. 
The final objective is:
\begin{equation}
\begin{aligned}
\mathcal{L}=\;&\sum_{k} w^{(k)}\,\big\lVert v_\theta(x_t^{(k)},t^{(k)})-v^{\star(k)}\big\rVert^2\\
&+\beta_{\mathrm{W2}}\sum_{k}\big\lVert v_\theta(x_t^{(k)},t^{(k)})-v_{\theta_{\mathrm{ref}}}(x_t^{(k)},t^{(k)})\big\rVert^2,
\end{aligned}
\label{eq:loss}
\end{equation}
where the first term is the advantage-weighted flow-matching loss, while the second uses the squared discrepancy between the policy and
reference velocities evaluated at the same noised input $x_t^{(k)}$ as a
tractable upper bound on their Wasserstein-2 distance.
% the second anchors the policy to the reference through a Wasserstein-2 penalty.
The signed weights encourage fitting above-average candidates while suppressing
below-average ones. 
However, negative weights can drive the corresponding squared regression errors to grow without bound, thereby destabilizing optimization.
The reference penalty mitigates this tendency by keeping the fine-tuned velocity
field close to the pretrained model, with $\beta_{\mathrm{W2}}$ controlling the
trade-off. Only the policy parameters $\theta$ receive gradients. The rollouts,
rewards, and reference velocities are all computed without gradients.
\begin{table*}[t]
\centering
{\small
\setlength{\tabcolsep}{1pt}
\begin{tabular}{l cc ccc ccc ccc ccc}
\toprule
\multirow{2}{*}{Method} & \multirow{2}{*}{NFE} & \multirow{2}{*}{Speed-up} & \multicolumn{3}{c}{LibriSpeech test-clean} & \multicolumn{3}{c}{Seed-TTS EN} & \multicolumn{3}{c}{Seed-TTS ZH} & \multicolumn{3}{c}{Avg.\ $\Delta$ vs.\ Pretrain} \\
\cmidrule(lr){4-6} \cmidrule(lr){7-9} \cmidrule(lr){10-12} \cmidrule(lr){13-15}
 & & & WER$\downarrow$ & SIM$\uparrow$ & UTMOS$\uparrow$ & WER$\downarrow$ & SIM$\uparrow$ & UTMOS$\uparrow$ & WER$\downarrow$ & SIM$\uparrow$ & UTMOS$\uparrow$ & WER$\uparrow$ & SIM$\uparrow$ & UTMOS$\uparrow$ \\
\midrule
Pretrain & -- & -- & 2.373 & 0.648 & 4.30 & 2.406 & 0.663 & 3.95 & 1.269 & 0.717 & 3.14 & -- & -- & -- \\
\midrule
DiTAR-GRPO & 32 & 1$\times$ & 2.149 & 0.674 & 4.29 & 2.017 & 0.688 & 3.91 & 1.144 & 0.736 & 2.97 & 0.246 & 0.023 & -0.07 \\
 & 10 & 2.6$\times$ & 2.332 & 0.683 & 4.31 & 1.728 & 0.699 & 3.97 & 1.180 & 0.738 & 2.98 & 0.269 & 0.030 & -0.04 \\
 & 5 & 4.0$\times$ & 2.393 & 0.665 & 4.21 & 1.873 & 0.688 & 3.93 & 1.194 & 0.730 & 2.91 & 0.196 & 0.018 & -0.11 \\
\midrule
\textbf{GROW} & 32 & 1.1$\times$ & \textbf{1.868} & \textbf{0.701} & \textbf{4.33} & 1.750 & \textbf{0.703} & 4.01 & 1.000 & \textbf{0.743} & 3.13 & \textbf{0.477} & \textbf{0.040} & 0.03 \\
 & 10 & 2.9$\times$ & 1.927 & \textbf{0.701} & \textbf{4.33} & 1.763 & 0.702 & 4.00 & \textbf{0.983} & 0.742 & 3.12 & 0.458 & 0.039 & 0.02 \\
 & 5 & 4.3$\times$ & 1.992 & 0.697 & \textbf{4.33} & \textbf{1.721} & 0.702 & \textbf{4.04} & 1.096 & 0.742 & \textbf{3.19} & 0.413 & 0.038 & \textbf{0.06} \\
\bottomrule
\end{tabular}}
% \caption{Zero-shot TTS performance of GROW compared with DiTAR-GRPO baseline, applied on the same pretrained DiTAR model and evaluated on three test sets. The ``Avg.\ $\Delta$ vs.\ Pretrain'' block reports the absolute improvement over the pretrained model, averaged over the three test sets (higher is better in every column). The best value in each column is set in \textbf{bold}. Effect of rollout NFE on the training efficiency and zero-shot
% TTS performance of GROW and DiTAR-GRPO. All models are evaluated with
% NFE${=}32$. ``Speed-up'' reports training wall-clock speed relative to
% DiTAR-GRPO with rollout NFE${=}32$.}
\caption{Zero-shot TTS performance and training efficiency of GROW compared with DiTAR-GRPO baseline, applied on the same pretrained DiTAR model and evaluated on three test sets. 
%Effect of rollout NFE on the training efficiency and zero-shot TTS performance of GROW and DiTAR-GRPO. 
We vary the rollout NFE during training, while fixing the inference NFE at $32$ for all models. We use NFE${=}10$ as the default setting for the main experiments.
% The rollout NFE is varied only during RL training, and NFE${=}10$ is the default setting used for the main comparison, while all models are evaluated with NFE${=}32$.
% All models are evaluated with NFE${=}32$. 
``Speed-up'' reports training wall-clock speed relative to
   DiTAR-GRPO with rollout NFE${=}32$. The ``Avg.\ $\Delta$ vs.\ Pretrain'' block reports the absolute improvement over the pretrained model, averaged over the three test sets (higher is better in every column). The best value in each column is set in \textbf{bold}.}

% \caption{Zero-shot TTS performance and training efficiency of GROW and
% DiTAR-GRPO under different rollout NFEs. The rollout NFE is varied only
% during RL training: NFE${=}10$ is the default setting used for the main
% comparison, while all final evaluations use NFE${=}32$. ``Speed-up''
% reports training wall-clock speed relative to DiTAR-GRPO with rollout
% NFE${=}32$. The ``Avg.\ $\Delta$ vs.\ Pretrain'' block reports the
% absolute improvement over the pretrained model, averaged across the
% three test sets (higher is better in every column). The best value in
% each column is shown in \textbf{bold}.}

\label{tab:efficiency}
\end{table*}
% \caption{Effect of rollout NFE on the training efficiency and zero-shot
% TTS performance of GROW and DiTAR-GRPO. All models are evaluated with
% NFE${=}32$. Speed-up is the rollout wall-clock speed relative to the FLOW-GRPO rollout-NFE${=}32$ setting (higher is faster).}
% %
% Effect of the on-policy rollout NFE used during RL training for GROW-CFM-W2 and FLOW-GRPO. We ablate the
% number of function evaluations (NFE) that the flow-matching sampler takes to generate each latent patch during on-policy
% rollout, and report both the resulting training speed and the final zero-shot TTS quality on the three test sets. The final evaluation
% always uses a fixed NFE of 32, so the NFE column reflects a training-time rollout choice only. Speed-up is the rollout wall-clock

\section{Experiments}
\subsubsection{Experimental Setup}

\paragraph{Datasets.}
% large-scale in-the-wild 
% filtered for transcription and language errors following the F5-TTS protocol
For pre-training, we use the English and Chinese portions of Emilia corpus~\citep{he2024emilia}, comprising $\sim$ 95K hours of speech. For RL, we assemble a balanced multilingual prompt set of 30K utterances drawn in equal parts from three sources: 10K English and 10K Chinese utterances sampled from Emilia, and 10K English utterances sampled from LibriTTS train-clean-100~\citep{zen2019libritts}. 
Emilia
utterances are retained only when their DNSMOS exceeds $3.2$. Every prompt is
constrained to a duration between $1.5$ and $15$ seconds. Each training instance couples a prompt waveform and its transcript with a target text sampled randomly from the same dataset.
%from the texts of the same dataset as the prompt waveform. 
The length of target text varies from short to long, spanning a range of synthesis difficulties.

%to enlarge the reward spread within a rollout group, $30\%$ of the target texts are drawn from the upper duration band above the $80$-th percentile. 
%The construction uses a fixed random seed for reproducibility.

\paragraph{Pre-training Configurations.}
We use Semantic-VAE~\citep{niu2025semantic} for continuous speech representation that encodes $16$~kHz waveforms into 40Hz $64$-dimensional latents. 
A six-layer transformer aggregation encoder then processes each patch of $P{=}4$ consecutive latents into a single vector, which is fed as one input token to the causal AR backbone.
% DiTAR groups every $P\!=\!4$ consecutive latents into a patch, which a six-layer aggregation encoder compresses into a single token for the autoregressive backbone. 
The backbone is a Qwen3-0.6B Transformer~\citep{yang2025qwen3} with $28$ layers and a hidden size of $1024$, 
%$16$ attention heads under grouped-query attention with $8$ key-value heads, and a feed-forward size of $3072$; 
initialized from the released weights.
%Qwen3-0.6B parameters. 
The input text is tokenized using Qwen3-0.6B tokenizer. The LocDiT is a six-layer transformer with a model dimension of $1024$ and 120M parameters. The number of historical patch is set to 1.
% The complete model comprises roughly $0.6$B parameters. 
% Training minimizes the linear flow-matching objective of Eq.~\eqref{eq:cfm} together with an auxiliary
% stop-prediction loss under equal loss weights
% stop 先不介绍了
The causal LM output is dropped with probability $0.1$ to enable CFG at sampling time. 
From the causal LM output, a lightweight linear stop head makes a binary continue-or-stop decision, trained with a cross-entropy loss.
We optimize with AdamW optimizer, and the learning rate increases linearly from 0 to a peak rate of 1e-4 over the first 10k steps, and then decays linearly to zero during the remaining training time. 
% using frame-based dynamic batching, a gradient-norm clip of $1.0$, and bf16 mixed precision. 
The checkpoint at $200$K updates serves as both the initialization and the frozen reference $q$ for the RL stage.

\paragraph{RL Fine-tuning Configurations.}
We fine-tune the $200$K-step pre-trained model with the objective of Eq.~\eqref{eq:loss}. 
% Both the policy and the frozen reference are initialized from this checkpoint, so their divergence is zero at the first update. 
Each optimization step draws a group of $G\!=\!8$ on-policy rollouts.
% every rollout is generated by solving the sampling ODE for $10$ steps. 
% Speaker similarity is scored by the cosine distance between WavLM-large ECAPA-TDNN speaker embeddings~\citep{}, and intelligibility by $1\!-\!\mathrm{WER}$ using
% faster-whisper large-v3~\citep{} for English and Paraformer-zh~\citep{} for Chinese;
% both reward models are frozen and are shared with evaluation. 
The two standardized reward streams of Eq.~\eqref{eq:advantage} are fused with equal weights $\lambda_{\mathrm{wer}}\!=\!\lambda_{\mathrm{sim}}\!=\!1$, and the W2
reference anchor of Eq.~\eqref{eq:loss} is weighted by $\beta_{\mathrm{W2}}\!=\!0.025$. 
%The Semantic-VAE generator is kept frozen throughout. 
% For the standardized reward streams of Eq.~\eqref{eq:advantage}, we set both fusion weights to $\lambda_{\mathrm{wer}}\!=\!\lambda_{\mathrm{sim}}\!=\!1$ and the reference anchor to $\beta\!=\!0.025$.
We use AdamW with a constant learning rate of $2\times10^{-6}$. The batch size is set to 128.
%gradients accumulated over two micro-steps, 
% a gradient-norm clip of $5.0$, and bf16 mixed precision; each micro-step consumes a single prompt and its group of rollouts. 
The RL stage runs for several hundred updates.
Both on-policy rollout and final evaluation use the same sampling procedure.  % numerically integrate the sampling ODE with an Euler solver, 
Specifically, we use an Euler solver, adopt the sway sampling schedule of F5-TTS with a coefficient of $-1$, and apply CFG. The only difference lies in the number of function evaluations(NFEs), which is $10$ during rollout and $32$ at evaluation. Note that the flow-matching regression loss in Eq.~\eqref{eq:loss} is computed using the conditional velocity prediction alone without CFG.

\paragraph{Evaluation Metrics.}
Following standard practice, we evaluate on the LibriSpeech-PC test-clean subset proposed in F5-TTS, including 1,127 audio clips, and the Seed-TTS test sets, comprising $1{,}088$ English and $2{,}020$ Chinese prompt--target pairs. We report WER for content consistency computed by Whisper-large-v3~\citep{radford2023robust} and Paraformer-zh~\citep{gao2022paraformer}, speaker similarity (SIM) for timbre fidelity via cosine similarity between the WavLM-large-based speaker embeddings~\citep{spksim} of the generated and prompt speech, and UTMOS~\citep{utmos} for perceptual quality. These are the same recognition and speaker models used to compute the RL reward.

% Synthesis uses the Euler ODE sampler with $32$ function evaluations and a sway-sampling coefficient of $-1$.

\subsubsection{Experimental Result}

\paragraph{Comparison with the DiTAR-GRPO baseline.}
% \paragraph baseline}
As a strong point of comparison, we implement GRPO on the same pretrained DiTAR model, which we denote as DiTAR-GRPO. %, following the open-source RL recipe of Flow-GRPO. 
% Because flow matching is deterministic, this line of work converts the sampling ODE into an equivalent
% SDE so that each denoising step becomes a Gaussian policy amenable to a
% group-relative policy gradient.   two recent refinements: the 
We further strengthen the baseline with a mixed ODE--SDE scheme
of MixGRPO~\citep{li2025mixgrpo}, which confines the injected stochasticity and
its gradients to a short window of three denoising steps for efficiency, and the
CPS method, which keeps the
per-step noise level consistent with the scheduler~\citep{wang2025cps}.

% 对比了grpo和grow，grpo有不少缺点 cumbersome，但这段适合写到前面。
% Even so, running GRPO on a flow-matching TTS model remains cumbersome for
% three reasons. First, the baseline must inject noise into the deterministic
% denoising steps just to make the model trainable, and this noise also distorts
% the very utterances that the reward models then score. Second, the method must
% track a log-probability and a divergence-to-reference term step by step along
% the trajectory, rather than optimizing a single regression loss. Third, it is
% delicate to tune. Too little injected noise leaves the model unable to explore,
% whereas too much noise ruins the samples, and the high cost of each rollout
% forces a small candidate group that makes the advantage estimate noisy.
% GROW avoids all of this: it acts directly on the deterministic generator,
% re-weighting the standard flow-matching regression by a group-relative
% advantage with a Wasserstein-2 anchor to the reference (Eq.~\eqref{eq:loss}),
% so the policy is trained on clean, on-distribution samples.

Table~\ref{tab:efficiency} reports zero-shot performance across the three test sets.
Both RL methods improve over the pretrained DiTAR.
% , confirming RL fine-tuning is effective on this backbone. 
GROW, however, improves by a substantially larger and more consistent margin. Averaged over the three test sets, GROW lowers WER by $0.458$ and
raises SIM by $0.039$, against $0.269$ and $0.030$ of DiTAR-GRPO.
%, roughly $70\%$ and $30\%$ larger, respectively. 
Crucially, GROW improves the two objectives simultaneously without trading one off
against the other. 
It attains the highest SIM on all three test sets
and the lowest WER on LibriSpeech-PC and Seed-TTS ZH.
% Crucially, these two gains are simultaneous rather than traded against each other. GROW attains the best SIM on all three test sets and the best WER on LibriSpeech-PC and Seed-TTS ZH, so it advances intelligibility and speaker fidelity at the same time. 
The gains are also consistent rather than concentrated on a single test set.
Whereas DiTAR-GRPO improves mostly on one English set, GROW improves on
all three, which indicates that the improvements are robust properties of the
method rather than dataset-specific artifacts. %its behavior is reassuring. 
Although UTMOS is reported only as a held-out perceptual sanity check and never
enters the reward, GROW improves it whereas DiTAR-GRPO degrades it on average. We attribute this
degradation to the noise injected by its SDE conversion, which our
deterministic, reward-weighted update sidesteps. All values nonetheless remain
within a normal range.

\begin{table}[t]
\centering
{\small
\setlength{\tabcolsep}{4pt}
\begin{tabular}{llcccc}
\toprule
\multirow{2}{*}{Test set}
& \multirow{2}{*}{Metric}
& \multirow{2}{*}{Pretrain}
& \multicolumn{3}{c}{Weight} \\
\cmidrule(lr){4-6}
& & &
$\exp(\tau A)$
& $\exp(\tau r)$
& $A/\sigma$ \\
\midrule

\multirow{2}{*}{LS-PC}
& WER$\downarrow$
& 2.373 & 2.223 & 2.361 & \textbf{1.927} \\
& SIM$\uparrow$
& 0.648 & 0.646 & 0.647 & \textbf{0.701} \\

\midrule
\multirow{2}{*}{Seed EN}
& WER$\downarrow$
& 2.406 & 2.102 & 2.251 & \textbf{1.763} \\
& SIM$\uparrow$
& 0.663 & 0.662 & 0.666 & \textbf{0.702} \\

\midrule
\multirow{2}{*}{Seed ZH}
& WER$\downarrow$
& 1.269 & 1.127 & 1.151 & \textbf{0.983} \\
& SIM$\uparrow$
& 0.717 & 0.712 & 0.716 & \textbf{0.742} \\

\midrule
\multirow{2}{*}{Avg.\ $\Delta$}
& WER$\uparrow$
& -- & 0.199 & 0.095 & \textbf{0.458} \\
& SIM$\uparrow$
& -- & -0.003 & 0.000 & \textbf{0.039} \\

\bottomrule
\end{tabular}}
\caption{Effect of the per-rollout weighting function on zero-shot TTS
performance. We compare the signed group-normalized advantage
$A(x)/\sigma$ with exponential weighting of the centered advantage
$\exp(\tau A(x))$ and raw reward $\exp(\tau r(x))$.
LS-PC denotes LibriSpeech test-clean.}
\label{tab:weight}
\end{table}

\begin{table}[t]
\centering
\small
\setlength{\tabcolsep}{3pt}
\begin{tabular}{llcccc}
\toprule
\multirow{2}{*}{Test set}
& \multirow{2}{*}{Metric}
& \multirow{2}{*}{Pretrain}
& \multicolumn{3}{c}{Configuration $(\gamma,v)$} \\
\cmidrule(lr){4-6}
& & &
$(1.5,v_c)$
& $(1.5,v_g)$
& $(0,v_c)$ \\
\midrule

\multirow{2}{*}{LS-PC}
& WER$\downarrow$
& 2.373 & \textbf{1.927} & 2.287 & 1.953 \\
& SIM$\uparrow$
& 0.648 & \textbf{0.701} & 0.657 & 0.688 \\

\midrule
\multirow{2}{*}{Seed EN}
& WER$\downarrow$
& 2.406 & 1.763 & 2.283 & \textbf{1.648} \\
& SIM$\uparrow$
& 0.663 & \textbf{0.702} & 0.671 & 0.695 \\

\midrule
\multirow{2}{*}{Seed ZH}
& WER$\downarrow$
& 1.269 & \textbf{0.983} & 1.195 & 1.052 \\
& SIM$\uparrow$
& 0.717 & 0.742 & 0.723 & \textbf{0.744} \\

\midrule
\multirow{2}{*}{Avg.\ $\Delta$}
& WER$\uparrow$
& -- & 0.458 & 0.094 & \textbf{0.465} \\
& SIM$\uparrow$
& -- & \textbf{0.039} & 0.007 & 0.033 \\

\bottomrule
\end{tabular}
\caption{Effect of rollout CFG and training velocity on zero-shot TTS
performance. We vary the rollout guidance scale $\gamma$ and apply the
flow-matching loss to either the conditional prediction $v_c$ or the
guided prediction $v_g$.}
\label{tab:cfg_loss}
\end{table}

\paragraph{Rollout NFE and training efficiency.}
Both RL methods spend most of each training step on the on-policy rollout.
Since DiTAR generates autoregressively, producing a single candidate runs a
full ODE for every latent patch. Each ODE is integrated over a
number of denoising steps that we treat as the rollout budget and report as the
\textit{NFE} column of Table~\ref{tab:efficiency}. 
Besides, the LocDiT forwards twice under CFG at every step, once conditionally and once unconditionally.
% flow-matching
% A rollout thus costs on the order of the patch count $\times$ NFE decoder computation,
% which makes it 
% the dominant term in the per-step
% wall-clock. $2 \times
A rollout thus requires on the order of
$\text{patch count}\times\text{NFE}$
LocDiT forward passes, making it the dominant contributor to the per-step wall-clock time.
Since Flow-GRPO shows coarse sampler suffices to
collect on-policy data even when fine sampling is needed at test time, we reduce
the rollout NFE in training while always evaluating the final model at
$\mathrm{NFE}=32$. Table~\ref{tab:efficiency} sweeps $\mathrm{NFE}\in\{32,10,5\}$ for both
methods.

For GROW, reducing the rollout NFE barely changes final quality. Cutting
NFE from $32$ to $10$ shrinks the average WER reduction by only $0.019$, from
$0.477$ to $0.458$, and cutting it to $5$ shrinks it by $0.064$, to $0.413$,
while speaker similarity stays essentially flat at $0.040$, $0.039$, and
$0.038$.%Since the rollout dominates the step, 
% These small quality changes translate into large speed-ups. 
This robustness to lower NFE enables substantial speed-ups
with only minor losses in quality.
Relative to the slowest configuration, DiTAR-GRPO
at NFE${=}32$, GROW trains $1.1\times$, $2.9\times$, and $4.3\times$ faster
at the three settings. We adopt NFE${=}10$ as the default for preserving
essentially the full quality of NFE${=}32$ while training $2.9\times$ faster.
DiTAR-GRPO tolerates a moderate reduction but breaks down under an aggressive one. Reducing its NFE to $10$ even improves both metrics
slightly at a $2.6\times$ speed-up, but at NFE${=}5$ its quality falls
clearly, despite a comparable $4.0\times$ speed-up.
%raising the WER reduction from $0.246$ to $0.269$ and the SIM gain from $0.023$ to $0.030$ 
% to a WER reduction of $0.196$ and a SIM gain of $0.018$

% At any fixed NFE the two methods incur the same rollout cost, since both
% draw the $8$ candidates with the identical sampler and CFG, so GROW's consistent $\sim\!1.1\times$ edge,
% $1.1/2.9/4.3\times$ against DiTAR-GRPO's $1.0/2.6/4.0\times$, comes entirely from
% the gradient propagation stage.
At matched NFE, the two methods use identical rollout configurations, so GROW's consistent $\sim\!1.1\times$ speed advantage over
DiTAR-GRPO comes entirely from the gradient-propagation stage.
%, where the two updates place their cost on different parts of the network. 
GROW performs a single flow-matching regression, one LocDiT
forward pass per frame, but backpropagates through the full AR backbone. DiTAR-GRPO utilizes the recorded LM output and backpropagates through the LocDiT alone, but 
% , yet it must replay the three SDE steps it injected per
% patch and rebuild the guided velocity at each with a conditional and an unconditional forward pass for both the policy and the reference, roughly $6\times$ as many decoder forward passes. 
the three SDE steps it makes stochastic per patch each enter the loss with a conditional and an unconditional forward pass, resulting in roughly $6\times$ more computation.
%The two costs thus move in opposite directions and neither update is dramatically cheaper than the other. 
% Both are moreover fixed by
% the patch and SDE-step counts and independent of NFE,  in any case 
However, the gradient stage is small compared to rollout stage that dominates the step, so the gap between the two loss computations is diluted to $\sim\!1.1\times$ as we observe. GROW is also lighter on GPU memory, as DiTAR-GRPO must additionally retain the recorded transition buffer for all rollout candidates.
Overall, GROW is faster, better, and more robust to
a shrinking rollout budget than DiTAR-GRPO.
%holding near its full-NFE quality even at NFE${=}5$ where the baseline degrades.

\paragraph{Linear versus exponential advantage weighting.}
How to map reward to the per-rollout weight $w(x)$ is a central
design choice in GROW. Table~\ref{tab:weight} ablates this mapping,
comparing our signed linear advantage $A(x)/\sigma$ against its two exponential ancestors
$\exp(\tau A(x))$ and $\exp(\tau r(x))$.
%  of~\eqref{eq:linear-adv}
% ~\eqref{eq:exp-adv}
% , which exponentiates the
% advantage, and $\exp(\tau r(x))$, the raw reward-weighted-regression weight that
% realizes the tilted optimum~\eqref{eq:opt-policy}.
The signed advantage is far more effective than either exponential weight.
Averaged over the three test sets, it lowers WER by $0.458$ and raises SIM by
$0.039$. The two exponential weights achieve a much smaller WER reduction, of
$0.199$ and $0.095$ respectively, and leave speaker similarity essentially
unmoved of $-0.003$ and $0$.
This is precisely the failure mode exposed by the
self-imitation/reward-contrast decomposition discussed above. 
The analysis applies to both weighting forms, for we implement $\exp(\tau r)$ with a group softmax, equivalent to the renormalized $\exp(\tau A)$. 
%of Section~\ref{sec:exp-to-linear}. Both exponential weights are non-negative, so their group mean $\bar w$ stays of order one while the reward-contrast deviations scale only as $O(\tau\sigma)$. 
%The weighted loss~\eqref{eq:decomp} is therefore 
The weighted loss is dominated by the self-imitation term, which reinforces the current
policy's own rollouts without ranking them. 
% Since a strong pretrained model
% keeps the within-group reward spread $\sigma$ small, 
The reward-contrast term becomes negligible, and the metrics barely move. 
% The residual signal yields a modest WER reduction but leaves the more saturated speaker-similarity reward at its starting point.
Enlarging $\tau$ is not a robust remedy, for $\sigma$ is unknown and drifting.
We swept the temperature $\tau$ together with
the W2 anchor weight $\beta_{\mathrm{W2}}$, and none of
the settings lifted speaker similarity above the pretrained model. 
The signed, zero-mean advantage with unit-variance normalization escapes this bind.
% The
% signed advantage escapes this bind by construction: being zero-mean, it cancels
% the self-imitation term and leaves an almost purely reward-contrast update;
% being signed, it pushes probability mass away from below-average rollouts rather
% than reinforcing every rollout to some degree; and its unit-variance
% normalization renders the signal scale independent of $\sigma$. 
We adopt $A(x)/\sigma$ as the default weight throughout.

\paragraph{Rollout CFG and training velocity.}
% Our GROW draws its on-policy rollouts under CFG but
% re-weights the ordinary flow-matching regression, whose target is the
% conditional velocity $v_c$ predicted by the model from the condition. We take
% $v_c$ as the default target for two practical reasons: it doesn't update the
% unconditional branch, which seems non-critical for reinforcement
% stage, and it is the lighter option, avoiding the extra unconditional forward
% pass that forming the guided velocity
% $v_g = v_c + \gamma\,(v_c - v_\varnothing)$ would require. ($v_\varnothing$ denotes the unconditional velocity obtained by zeroing the context.) This still leaves two
% design choices worth probing, both ablated in Table~\ref{tab:cfg_loss}: whether
% the rollouts are drawn with CFG,
% and whether the loss regresses $v_c$ or the guided velocity $v_g$ actually used
% to draw the samples.
GROW uses CFG for on-policy rollout, sampling with the guided velocity
$v_g=v_c+\gamma(v_c-v_\varnothing)$, where $v_\varnothing$ denotes
the unconditional prediction obtained by zeroing the context. 
During training, we apply the advantage-weighted flow-matching loss to the conditional
prediction $v_c$ rather than $v_g$. This choice focuses the RL update on
conditional generation, %which is directly evaluated by the rewards,
without explicitly optimizing the auxiliary unconditional prediction
used only for CFG. It also avoids the additional unconditional forward
pass required to construct $v_g$ in the loss.
Table~\ref{tab:cfg_loss} probes two alternative choices:
whether CFG is used during rollout, and whether the regression loss is
applied to the conditional prediction $v_c$ or the guided prediction
$v_g$.

Counterintuitively, switching from $v_c$ to $v_g$ makes the method
nearly ineffective: the average WER reduction falls from $0.458$ to
$0.094$ and the SIM gain from $0.039$ to $0.007$, leaving only marginal
improvements in both metrics across all test sets.
% , roughly a fifth of each
We therefore use conditional prediction $v_c$ in loss throughout our experiments.
With this choice fixed, the rollout guidance scale $\gamma$ shifts the balance between intelligibility and speaker fidelity. Disabling CFG at $\gamma{=}0$ so that rollouts
are drawn from pure conditional distribution yields a slightly larger WER reduction ($0.465$ vs.\ $0.458$) but a smaller  SIM gain
($0.033$ vs.\ $0.039$). Guiding the rollout at $\gamma{=}1.5$ instead
concentrates the candidates toward the conditioning, sharpening the speaker
adherence that SIM measures, and gives the best speaker similarity. 
%The two settings are thus a clean Pareto pair rather than a winner and a loser. 
We finally adopt $\gamma{=}1.5$, as it maximizes speaker similarity while preserving intelligibility, whereas $\gamma{=}0$ is an equally attractive efficiency-oriented alternative that additionally halves the rollout cost, skipping the unconditional forward pass at every denoising step.   

\begin{table}[t]
\centering
{\small
\setlength{\tabcolsep}{1pt}
\begin{tabular}{cccccccccc}
\toprule
\multirow{2}{*}{$\eta$}
& \multirow{2}{*}{$\beta_{\mathrm{W2}}$}
& \multicolumn{2}{c}{LS-PC}
& \multicolumn{2}{c}{Seed EN}
& \multicolumn{2}{c}{Seed ZH}
& \multicolumn{2}{c}{Avg.\ $\Delta$} \\
\cmidrule(lr){3-4}
\cmidrule(lr){5-6}
\cmidrule(lr){7-8}
\cmidrule(lr){9-10}
& & WER$\downarrow$ & SIM$\uparrow$
& WER$\downarrow$ & SIM$\uparrow$
& WER$\downarrow$ & SIM$\uparrow$
& WER$\uparrow$ & SIM$\uparrow$ \\
\midrule

\multirow{5}{*}{$2\mathrm{e}^{-6}$}
& 0
& 2.31 & 0.64
& 2.23 & 0.66
& 1.11 & 0.71
& 0.134 & -0.003 \\

& 0.025
& \textbf{1.93} & \textbf{0.70}
& \textbf{1.76} & \textbf{0.70}
& \textbf{0.98} & \textbf{0.74}
& \textbf{0.458} & \textbf{0.039} \\

& 0.05
& 2.13 & 0.69
& 1.89 & 0.69
& 0.99 & \textbf{0.74}
& 0.344 & 0.030 \\

& 0.1
& 1.98 & 0.67
& 1.84 & 0.69
& 1.15 & 0.73
& 0.360 & 0.020 \\

& 1
& 2.37 & 0.65
& 2.28 & 0.67
& 1.19 & 0.72
& 0.068 & 0.004 \\

\midrule

$1\mathrm{e}^{-6}$
& 0.025
& 1.98 & 0.69
& 1.87 & \textbf{0.70}
& 1.11 & \textbf{0.74}
& 0.362 & 0.035 \\

$5\mathrm{e}^{-6}$
& 0.1
& 2.14 & 0.67
& 2.03 & 0.68
& 1.13 & 0.73
& 0.251 & 0.017 \\

$1\mathrm{e}^{-5}$
& 0.1
& 2.30 & 0.67
& 2.00 & 0.68
& 1.05 & 0.73
& 0.231 & 0.018 \\

\bottomrule
\end{tabular}}
\caption{Effect of the learning rate $\eta$ and W2-anchor weight
$\beta_{\mathrm{W2}}$ on zero-shot TTS performance. }
\label{tab:lr_kl}
\end{table}

% Bold values are determined from the unrounded results.
% \begin{table*}[t]
% \centering
% {\small
% \setlength{\tabcolsep}{4pt}
% \begin{tabular}{c ccc ccc ccc ccc}
% \toprule
% \multirow{2}{*}{Reward} & \multicolumn{3}{c}{LibriSpeech-PC test-clean} & \multicolumn{3}{c}{Seed-TTS EN} & \multicolumn{3}{c}{Seed-TTS ZH} & \multicolumn{3}{c}{Avg.\ $\Delta$ vs.\ Pretrain} \\
% \cmidrule(lr){2-4} \cmidrule(lr){5-7} \cmidrule(lr){8-10} \cmidrule(lr){11-13}
%  & WER$\downarrow$ & SIM$\uparrow$ & UTMOS$\uparrow$ & WER$\downarrow$ & SIM$\uparrow$ & UTMOS$\uparrow$ & WER$\downarrow$ & SIM$\uparrow$ & UTMOS$\uparrow$ & WER$\uparrow$ & SIM$\uparrow$ & UTMOS$\uparrow$ \\
% \midrule
% WER & 1.958 & 0.649 & \textbf{4.33} & \textbf{1.646} & 0.663 & \textbf{4.00} & 1.087 & 0.720 & 3.12 & 0.452 & 0.001 & 0.02 \\
% SIM & 2.291 & \textbf{0.704} & \textbf{4.33} & 1.963 & \textbf{0.706} & 3.99 & 1.047 & \textbf{0.748} & \textbf{3.16} & 0.249 & \textbf{0.043} & \textbf{0.03} \\
% WER${+}$SIM & \textbf{1.927} & 0.701 & \textbf{4.33} & 1.763 & 0.703 & \textbf{4.00} & \textbf{0.983} & 0.743 & 3.12 & \textbf{0.458} & 0.039 & 0.02 \\
% \bottomrule
% \end{tabular}}
% \caption{Effect of reward composition on zero-shot TTS performance.
% We compare intelligibility-only, speaker-similarity-only, and combined
% rewards. }
% \label{tab:reward}
% \end{table*}

\begin{table}[t]
\centering
\small
\setlength{\tabcolsep}{3pt}
\begin{tabular}{llcccc}
\toprule
\multirow{2}{*}{Test set}
& \multirow{2}{*}{Metric}
& \multirow{2}{*}{Pretrain}
& \multicolumn{3}{c}{Reward} \\
\cmidrule(lr){4-6}
& & &
WER
& SIM
& WER+SIM \\
\midrule

\multirow{2}{*}{LS-PC}
& WER$\downarrow$
& 2.373 & 1.958 & 2.291 & \textbf{1.927} \\
& SIM$\uparrow$
& 0.648 & 0.649 & \textbf{0.704} & 0.701 \\

\midrule
\multirow{2}{*}{Seed EN}
& WER$\downarrow$
& 2.406 & \textbf{1.646} & 1.963 & 1.763 \\
& SIM$\uparrow$
& 0.663 & 0.663 & \textbf{0.706} & 0.702 \\

\midrule
\multirow{2}{*}{Seed ZH}
& WER$\downarrow$
& 1.269 & 1.087 & 1.047 & \textbf{0.983} \\
& SIM$\uparrow$
& 0.717 & 0.720 & \textbf{0.747} & 0.742 \\

\midrule
\multirow{2}{*}{Avg.\ $\Delta$}
& WER$\uparrow$
& -- & 0.452 & 0.249 & \textbf{0.458} \\
& SIM$\uparrow$
& -- & 0.001 & \textbf{0.043} & 0.039 \\

\bottomrule
\end{tabular}
\caption{Effect of reward composition on zero-shot TTS performance.
We compare intelligibility-only, speaker-similarity-only, and combined
rewards.}
\label{tab:reward}
\end{table}

% \caption{Ablation on the reward used to drive optimisation in DiTAR-ORW, comparing a WER-only reward, a SIM-only reward, and the combined WER${+}$SIM reward. We report zero-shot TTS performance for each configuration applied on top of the same pretrained diffusion model and evaluated on three test sets. 
% }

\paragraph{Learning rate and W2-anchor strength.}
Two important knobs in GROW are the learning rate $\eta$ and the reference-anchor weight $\beta_{\mathrm{W2}}$, which controls how tightly the policy is tethered to the frozen pretrained model. 
With $\eta$ fixed at $2\mathrm{e}{-}6$, the overall performance is best at an intermediate anchor strength.
% first improves as $\beta_{\mathrm{W2}}$ grows and then declines once the anchor becomes too strong, so it peaks at an intermediate value. 
Removing the anchor at
$\beta_{\mathrm{W2}}{=}0$ yields only modest WER gains and is the only
configuration that fails to improve speaker similarity, suggesting that the unchecked
signed advantage drifts off the pretrained manifold rather than concentrating on
high-reward behavior. Conversely, an over-strong anchor at $\beta_{\mathrm{W2}}{=}1$ instead pins the policy so tightly that it barely moves, sharply reducing the gains on
both metrics. The optimum lies in between at
$\beta_{\mathrm{W2}}{=}0.025$, which attains both the largest WER reduction and 
the largest SIM gain, while the intermediate values $0.05$ and $0.1$ form 
a similar but clearly lower plateau. 
This confirms that GROW benefits from a light but nonzero reference
anchor, which stabilizes training without impeding policy improvement.

% The reference anchor is therefore necessary,
% since without it speaker fidelity degrades, yet it must stay light so that the
% reward signal retains enough room to move the model.

%  of Eq.~\eqref{eq:loss}
% Table~\ref{tab:lr_kl} ablates both.
% This ablation tests the design argument of Section~\ref{sec:exp-to-linear},
% where the signed advantage weight is inherently divergent under a squared
% flow-matching loss and is contained jointly by the reference anchor and a small
% learning rate. Each knob turns out to have a pronounced interior optimum.

% The learning rate behaves similarly.
% Since each row in the lower block of
% Table~\ref{tab:lr_kl} carries its own $\beta_{\mathrm{W2}}$, we read the
% learning-rate effect only within a fixed anchor strength. 
At the optimal anchor
$\beta_{\mathrm{W2}}{=}0.025$, halving the learning rate to $1\mathrm{e}{-}6$ preserves training stability but yields smaller gains, indicating the updates are too conservative
to move the model sufficiently.
At the looser anchor $\beta_{\mathrm{W2}}{=}0.1$, raising the learning
rate to $5\mathrm{e}{-}6$ and then $1\mathrm{e}{-}5$ steadily shrinks the
WER reduction, indicating that overly aggressive updates move the model beyond
the neighborhood preserved by the reference anchor. The useful learning rate is
thus an intermediate value that requires tuning.
% confirming that the reward-weighted update helps only when kept close to the reference. 
Taken together, the two optima confirm the signed advantage weight
would diverge on its own under the squared flow-matching loss and is kept stable
by the reference anchor and a small learning rate.
%  the analysis of Section~\ref{sec:exp-to-linear},  where

\paragraph{Reward design.}
% of Eq.~\eqref{eq:advantage}
We next isolate the contribution of the two reward streams by training with three reward configurations:
intelligibility alone, speaker similarity alone, and the two combined.
Table~\ref{tab:reward} reports the result.
Each single-objective reward produces a substantial gain on its own axis without
sacrificing the other. And the coupling between the two axes is not symmetric.
Rewarding intelligibility alone yields a large
WER reduction of 0.452, while holding speaker similarity at the pretrained level.
Rewarding speaker similarity alone yields a marked similarity gain of 0.043, and also reduces WER by a wide margin of 0.249, which is more than half of the reduction achieved by the intelligibility reward.
% Neither gain is bought at the expense of the other. 
%while holding speaker similarity at the pretrained level
% The  Rewarding speaker similarity alone 
%whereas rewarding intelligibility alone leaves speaker similarity essentially unchanged. 
A likely reason is that steering generations toward the
prompt speaker also makes them cleaner and easier to recognize, while the
reverse coupling is weak. Combining the two rewards then delivers both gains at
once. The joint reward matches the WER-only setting on intelligibility and keeps
almost all of the SIM-only setting gains on speaker similarity, reaching essentially the same level as the strong single-objective run on each axis simultaneously.

\section{Conclusions}
We present GROW, an on-policy RL method that fine-tunes AR-diffusion TTS by reweighting standard flow-matching regression with a signed, group-normalized advantage and stabilizing the update with a Wasserstein-2 reference anchor. Our analysis and ablations identify the weighting rule as the central design choice. When rewards from a strong pretrained model are tightly clustered, positive exponential weights are dominated by common self-imitation, whereas the zero-mean signed advantage preserves within-group reward contrast and supports joint optimization of intelligibility and speaker similarity. Across three multilingual zero-shot benchmarks, GROW consistently improves both WER and speaker similarity, preserves perceptual quality, and outperforms a DiTAR-GRPO baseline in both effectiveness and training efficiency. 
% Its performance also remains robust when the rollout NFE is substantially reduced. 
These results establish direct advantage-weighted flow matching as a concise and effective alternative to trajectory-level policy gradients for flow-matching TTS generators. 
% Because GROW requires only the standard flow-matching objective and black-box utterance-level rewards, extending it beyond DiTAR to other flow-matching speech models is a promising direction.

\bibliography{aaai2027}

% Check whether the conference requires a reproducibility checklist to be included in the paper.
% If so, you can uncomment the following line and ajust the path to include it.
% \input{ReproducibilityChecklist.tex}

\end{document}